\documentclass{aa}
\usepackage{graphicx}
\usepackage{natbib}
\bibpunct{(}{)}{;}{a}{}{,} 

\begin{document}

\title{X-rays from the jet in \object{3C\,273}:\\ clues from the radio--optical
spectra}

\titlerunning{The radio--optical--X-ray spectrum of 3C\,273's jet}

\author{S.~Jester \inst{1}
   \and H.-J.~R\"oser \inst{1}
   \and K.~Meisenheimer \inst{1}
   \and R.~Perley \inst{2}
}

\offprints{S.\,Jester, \email{jester@mpia.de}}
\date{Received 14 February 2002, in original form 17 January 2002 /
Accepted 22 February 2002}
\institute{Max-Planck-Institut f\"ur Astronomie, K\"onigstuhl~17,
   69117~Heidelberg, Germany \and NRAO, P.\,O.~Box 0, Socorro, NM~87801, USA}

\abstract{Using new deep VLA and HST observations of the large-scale
jet in 3C\,273 matched to 0\farcs3 resolution, we have detected excess
near-ultraviolet emission ($\lambda\,300\mbox{${\rm\ts\,nm}$}$) above a synchrotron
cutoff spectrum accounting for the emission from radio through optical
($\lambda\lambda 3.6\mbox{${\rm\ts cm}$}$--$620\mbox{${\rm\ts\,nm}$}$).  This necessitates a
two-component model for the emission.  The radio--optical--X-ray
spectral energy distributions suggest a common origin for the UV
excess and the X-rays from the jet. \keywords{Galaxies: jets --
quasars: individual: 3C\,273 -- radiation mechanisms: non-thermal}}

\maketitle
\section{Introduction}\label{s:intro}
Recent Chandra observations have shown that X-rays from large-scale
extragalactic jets are more common than expected from earlier
observations using the \emph{Einstein Observatory} and ROSAT
\citep{HK02}.  They are most likely of non-thermal origin, but
for many cases is has been impossible to establish the relative
contributions from synchrotron and inverse Compton emission.

One of these is the jet of the quasar 3C\,273.  Due to its large
angular extent of over 20\arcsec ($\approx$ 60\,kpc at
$z$=0.158\footnote{$H_0=70\,\mathrm{km\,s}^{-1} \mathrm{Mpc}^{-1}$;
$\Omega_0 = 0.3, \Omega_\Lambda=0.7$}) and its high surface brightness
in the radio, optical and in X-rays, it is readily observable and
serves as a sample case for the study of extragalactic jets.  The high
degree of optical polarisation \citep[10\%--20\%;][]{RM91} and the
overall agreement with the radio polarisation properties \citep{jetIV}
imply that the bulk of the jet's optical light is synchrotron emission
from the same electron population responsible for the radio emission.
However, the X-rays could be synchrotron emission from this same
electron population only for regions~A and B of the jet
\citep{Roe00,Mareta01}.  Since self-Compton scattering of the
radio-optical emission can account only for a small fraction of the
observed X-ray fluxes, \citet{Roe00} suggested to consider the X-rays
as synchrotron emission from a second, higher-energy electron
population.  As an alternative, a contribution by beamed inverse
Compton scattering of the cosmic microwave background has been
suggested \citep{Mareta01,Sam01}. A distinction between these models
is not possible based on the currently available X-ray data. Both
require somewhat exotic physical conditions. In this letter, we
present new radio-optical observations which suggest a common origin
for the X-rays and part of the jet's ultraviolet emission.

We employ observations at wavelengths of 3.6\mbox{${\rm\ts cm}$},
2.0\mbox{${\rm\ts cm}$}, 1.3\mbox{${\rm\ts cm}$}\ (VLA\footnote{The
National Radio Astronomy Observatory is a facility of the National
Science Foundation operated under cooperative agreement by Associated
Universities, Inc.}), 1.6\hbox{\,$\mu$m}\ (HST-NICMOS\footnote{Based
on observations made with the NASA/ESA Hubble Space Telescope,
obtained at the Space Telescope Science Institute, which is operated
by the Association of Universities for Research in Astronomy, Inc.\
under NASA contract No.~NAS5-26555.  These observations are associated
with proposals \#5980 and \#7848.}), 620\mbox{${\rm\ts\,nm}$}\ and
300\mbox{${\rm\ts\,nm}$}\ (HST-WFPC2\footnotemark[3]).  The WFPC2 data
have been described in \citet{Jes01}. Details of the NICMOS data are
contained in \citet{JesterDiss} and will be reported elsewhere
\citetext{Jester \emph{et al., in prep.}}, as will the full VLA data
set \citetext{Perley \emph{et al., in prep.}}.

\section{Observed spectra}\label{s:obsspec}

\begin{figure}
  \resizebox{\hsize}{!}{\includegraphics[origin=B,angle=270,clip]{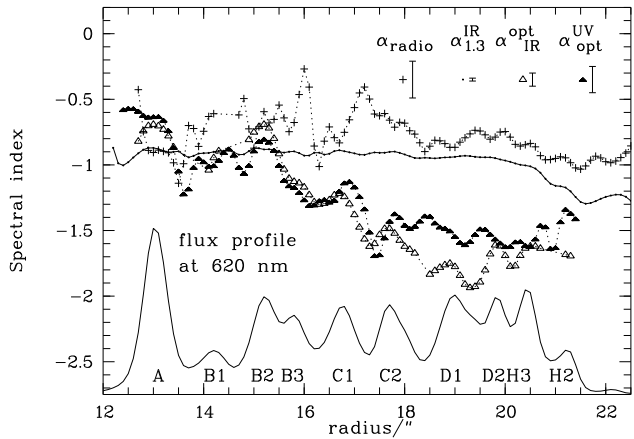}}
  \caption{Run of the spectral indices along the jet (0\farcs3
  resolution sampled at 0\farcs1 intervals). We show a cut along the
  radius vector at position angle 222\fdg2. For sake of clarity, only
  typical $2\sigma$ error bars are shown for the random error.
  Systematic flux calibration uncertainties are of the same order and
  would shift an entire curve. The radio spectral index
  $\alpha_\mathrm{radio}$ is obtained by a fit to the radio data at
  3.6\mbox{${\rm\ts cm}$}, 2.0\mbox{${\rm\ts cm}$}, and
  1.3\mbox{${\rm\ts cm}$}.  Other spectral indices are derived
  directly for the given wavelengths ($\alpha_{1.3}^{\mathrm{IR}}$:
  1.3\mbox{${\rm\ts cm}$}\ and 1.6\hbox{\,$\mu$m},
  \mbox{$\alpha^\mathrm{opt}_\mathrm{IR}$}: 1.6\hbox{\,$\mu$m}\ and
  620\mbox{${\rm\ts\,nm}$}, \mbox{$\alpha^\mathrm{UV}_\mathrm{opt}$}:
  620\mbox{${\rm\ts\,nm}$}\ and
  300\mbox{${\rm\ts\,nm}$}).\label{f:alpha-run}}
\end{figure}
We have performed photometry for circular Gaussian beams at common
FWHM of 0\farcs3.  We make sure that the relative alignment of
photometric apertures across all wavelengths is better than 0\farcs03,
so that identical volumes are sampled.  We first consider the observed
shape of the spectral energy distribution before describing it with a
theoretical model.

\begin{figure}
 \resizebox{\hsize}{!}{\includegraphics[clip]{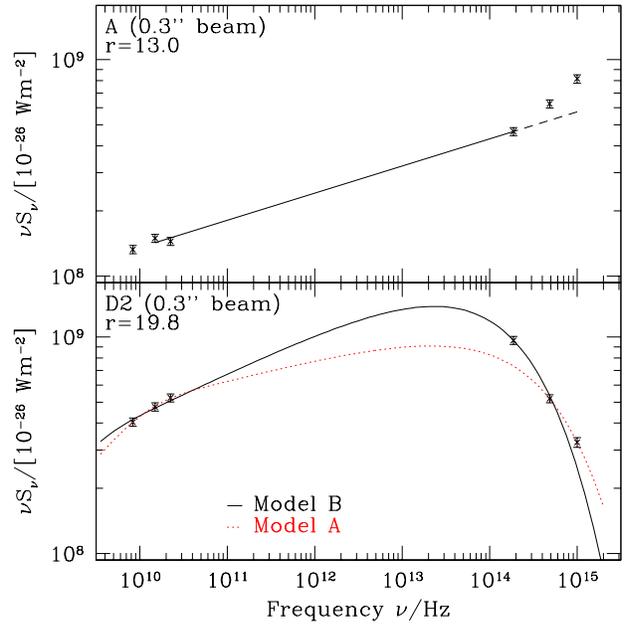}}
 \caption{Illustration of the two different spectral fits
 performed. Model A assumes infrared emission in excess of the cutoff
 given by the optical-ultraviolet spectral index, while Model B assumes
 an ultraviolet excess above the infrared-optical cutoff. No cutoff is
 observed for region A (upper panel).}
 \label{f:fitab}
\end{figure}

Figure\,\ref{f:alpha-run} shows the run of spectral indices along the
optically bright part of the jet (setting on at $r=12\arcsec$ from the
quasar core).  It shows remarkable features.  Firstly, variations in
$\alpha_{1.3}^{\mathrm{IR}}$ are much smaller than in other indices,
\mbox{\emph{i.\,e., }} the overall spectral shape is amazingly constant within the entire
jet (apart from the radio hot spot).  This reflects the jet's similar
appearance over more than four decades in frequency.  It implies that
there is a physical mechanism maintaining a similar electron
distribution everywhere in the jet.  Secondly, in regions A ($r\approx
13\arcsec$) and B2 ($r\approx 15\farcs2$), both the
ultraviolet-optical spectral index \mbox{$\alpha^\mathrm{UV}_\mathrm{opt}$}\ and the infrared-optical
index \mbox{$\alpha^\mathrm{opt}_\mathrm{IR}$}\ are flatter, \mbox{\emph{i.\,e., }} $|\alpha|$ is smaller, than the
radio-infrared spectral index $\alpha_{1.3}^{\mathrm{IR}}$ (see also
Fig.\,\ref{f:fitab}).  In regions C2 to D2/H3, from $r\approx
18\arcsec$ to $r\approx 20\arcsec$, \mbox{$\alpha^\mathrm{UV}_\mathrm{opt}$}\ is flatter than \mbox{$\alpha^\mathrm{opt}_\mathrm{IR}$},
although both are now steeper than the radio-infrared spectral index.
A similar flattening was already seen between \mbox{$\alpha^\mathrm{UV}_\mathrm{opt}$}\ determined with
the HST and the optical spectral index determined on ground-based data
at slightly lower frequencies \citep[Fig.\,4 in ][]{Jes01}.

To assess the significance of this flattening, it is necessary to
distinguish between various error sources.  The shot noise is less
than 0.5\% per 0\farcs3 beam in all bands.  The HST data have
flat-field errors of 1\% (WFPC2) -- 3\% \citep[NICMOS; compare the
error discussion in][]{Per01}.  The uncertainty in the background
estimation is estimated from the scatter in blank sky regions as
0.01\,$\mu$Jy\,per 0\farcs3 beam for WFPC2, and 0.03\,$\mu$Jy for
NICMOS, forming an error floor significant only for the faintest parts
of the jet.  An error source unique to the interferometric radio data
is the image fidelity, \mbox{\emph{i.\,e., }} errors in the sense that
the inferred brightness distribution does not correspond to the true
distribution on sky, in particular for the fainter parts of the jet.
We use a 3\% error to account for this.  All these error sources limit
the accuracy of relative photometry within one waveband, and the error
bars in Fig.\,\ref{f:alpha-run} show these errors.  In addition, all
wavebands will suffer an error from the absolute photometric
calibration, typically 2\%.  This calibration error changes all
measurements within one filter by the same factor, \mbox{\emph{i.\,e.,
}} it is a systematic error.  Its effect is to offset an entire
spectral index run by a constant amount.  The absolute photometry
errors are too small to allow matching the run of
\mbox{$\alpha^\mathrm{UV}_\mathrm{opt}$}\ with that of
\mbox{$\alpha^\mathrm{opt}_\mathrm{IR}$}\ in C2--H3, which in any case
would not remove the flattening of
\mbox{$\alpha^\mathrm{opt}_\mathrm{IR}$}\ below
$\alpha_{1.3}^{\mathrm{IR}}$ in A and B2.  The observed flattening is
thus real.

Observed synchrotron spectra are commonly described as power laws, and
are thought to arise from a power-law electron energy distribution.
Since synchrotron losses increase with the square of the electron
energy, any non-ideal electron distribution will give rise to a
spectrum with convex shape in $\log S_\nu$ against $\log\nu$,
\mbox{\emph{i.\,e., }} higher-frequency spectral indices are steeper
than those at lower frequencies, and any high-frequency flux must lie
below a power-law extrapolation from lower frequencies.  The observed
high-frequency flattening in A, B2, and C2--D2/H3 excludes that the
jet emission can be described using a simple, single electron
population.  Instead, a two-component model is required. The radio and
optical fluxes are dominated by the first component. A second
component must contribute in excess of this, either mainly in the
infrared, or mainly in the ultraviolet and at higher frequencies.

\section{Is there an infrared or ultraviolet excess?}\label{s:IR-or-X}
We describe the lower-frequency component by fitting model spectra
according to \citet{HM87}, which have been applied successfully to
observations of hot spots \citep{hs_II}.  These spectra
self-consistently account for synchrotron losses arising from a
Fermi-accelerated electron population which is continuously injected
into a loss region, and include losses even during the acceleration
phase. The resulting finite maximum particle energy leads to a
quasi-exponential cutoff at the synchrotron frequency corresponding to
the highest electron energies. These spectra are appropriate here as
the optically radiating electrons must be accelerated within the jet
itself \citep{Jes01}.

The original aim of these fits is the determination of the synchrotron
cutoff frequency and hence the maximum particle energy along the jet;
a detailed description and results will be published separately
\citep[Jester \emph{et al., in prep.}, and ][]{JesterDiss}. Here, we
consider the excess of the observed infrared or ultraviolet emission
above the spectrum arising from the lower-energy electron population,
and its implications for the X-ray emission mechanism.

We fit the observed radio-optical spectral energy distribution at
0\farcs3 beam size for all locations of the optical jet, performing
two separate fits which differ in the determination of the cutoff
frequency (Fig.\,\ref{f:fitab}): either the cutoff is described by the
optical-ultraviolet spectral index leading to an infrared excess
(Model A), or conversely, the true cutoff is described by the
infrared-optical spectrum and there is additional flux in the
ultraviolet (Model B).

Model A disregards the near-infrared flux point in the data set. This
is motivated by the indication that the radio cocoon around the jet
\citep{jetIV} may also be detectable at 2.1\hbox{\,$\mu$m}\ \citep{NMR97},
suggesting that the flux from the jet at 1.6\hbox{\,$\mu$m}\ may be
contaminated by emission from the cocoon as well. However, the
residuals in Model A contribute up to 30\% of the observed infrared
flux, which is a much larger fraction than expected for a cocoon
spectral energy distribution peaking at radio frequencies. Moreover, a
comparison of the residuals with our new high-resolution radio
polarisation map \citetext{Perley \emph{et al., in prep.}} shows that
the largest residuals are not cospatial with those regions showing the
highest fractional polarisation \citep{JesterDiss}. These properties
make this component very different from the original cocoon, which has
a steep spectrum (hence low flux at the short wavelengths considered
here) and high polarisation. In this model, the upward curvature of
the observed spectra at the inner end of the optical jet, in regions
A--B, remains unexplained. There is no plausible alternative infrared
emission process or source. Model A is therefore discarded.

Model B assigns a low weight to the ultraviolet flux point, so that
the location of the cutoff is dominated by the infrared and optical
points at 1.6\hbox{\,$\mu$m}\ and 620\mbox{${\rm\ts\,nm}$},
respectively. There is a significant ultraviolet excess above the
fitted infrared-optical cutoff or the radio-infrared power-law
extrapolation in all parts of the jet. As the actual cutoff could be
steeper than the fit including the UV point, the ultraviolet excess
determined in this way is actually a lower limit to the true
excess. At 300\mbox{${\rm\ts\,nm}$}, already a significant fraction of
the observed flux may be contributed by this additional component. A
smaller fraction of the optical emission will be contributed by the
same component. Unlike the cocoon, which would be expected to
contribute only in a limited part of the jet, the presumed additional
ultraviolet component may be present in the entire jet and can account
for the discrepancy between infrared-optical and optical-ultraviolet
spectrum everywhere it is observed.

As is the case for the X-ray emission, the origin of this excess
ultraviolet flux is unknown. Instead of postulating yet another
emission component, we suggest a common non-thermal origin for the
ultraviolet excess and the X-ray emission. To assess quantitatively
whether both could stem from the same electron population, we consider
whether the emission in both wavelength regions can be explained by a
single simple model.

\section{Discussion: The X-ray emission from the jet}\label{s:X-rays}
\begin{table}
\begin{center}
\begin{tabular}{lll}
\hline\hline
\multicolumn{1}{c}{Frequency}  & \multicolumn{2}{c}{Integrated flux
density [$\mu$Jy]} \\ 
\multicolumn{1}{c}{[Hz]} & Knot A & D2+H3\\
\hline
8.33 $\times 10^{9}$ & (78.7 $\pm$ 2.4) $\times 10^3$ & (866 $\pm$ 26) $\times 10^3$ \\
1.45 $\times 10^{10}$ & (50.4 $\pm$ 1.5) $\times 10^3$ & (536 $\pm$ 16) $\times 10^3$ \\
2.25 $\times 10^{10}$ & (35.5 $\pm$ 1.1) $\times 10^3$ & (371 $\pm$ 11) $\times 10^3$ \\
1.87 $\times 10^{14}$ & 9.66 $\pm$ 0.35  & 41.9 $\pm$ 1.5\\
4.84 $\times 10^{14}$ & 4.20 $\pm$ 0.09  &  7.56 $\pm$ 0.17\\
1.00 $\times 10^{15}$ & 2.40 $\pm$ 0.05  & 2.27 $\pm$ 0.05\\
2.42 $\times 10^{17}$ & 0.038 $\pm$ 0.004 & 0.0083 $\pm$ 0.0008\\
\hline\hline
$\alpha_\mathrm{X}$ & $-0.6 \pm .05$ & $-0.75 \pm .05$ \\
\hline
$L($radio) [W]  & 1.7 $\times 10^{36}$ & 1.0 $\times 10^{37}$ \\
$L($X-ray) [W] & 1.4 $\times 10^{36}$ & 4.1 $\times 10^{35}$ \\
\hline\hline
\end{tabular}
\end{center}
\caption{\label{t:sed} Emission from two regions of 3C\,273's jet. The
radio through optical data are from our new study.  The X-ray
flux and spectral index $\alpha_\mathrm{X}$ are taken from
\protect\citet{Mareta01}. $L($radio) is the luminosity in the lower-energy
component, assuming the radio spectrum below 5\,GHz continues to
10\,MHz with spectral index $\approx -0.5$. $L($X-ray) is the
UV--X-ray luminosity from $10^{15}$\,Hz to $2.42\times10^{17}$\,Hz at
the spectral index $\alpha_\mathrm{X}$.}
\end{table}
\begin{figure*}
\centering
\hfill\resizebox{.46\hsize}{!}{\includegraphics[angle=270]{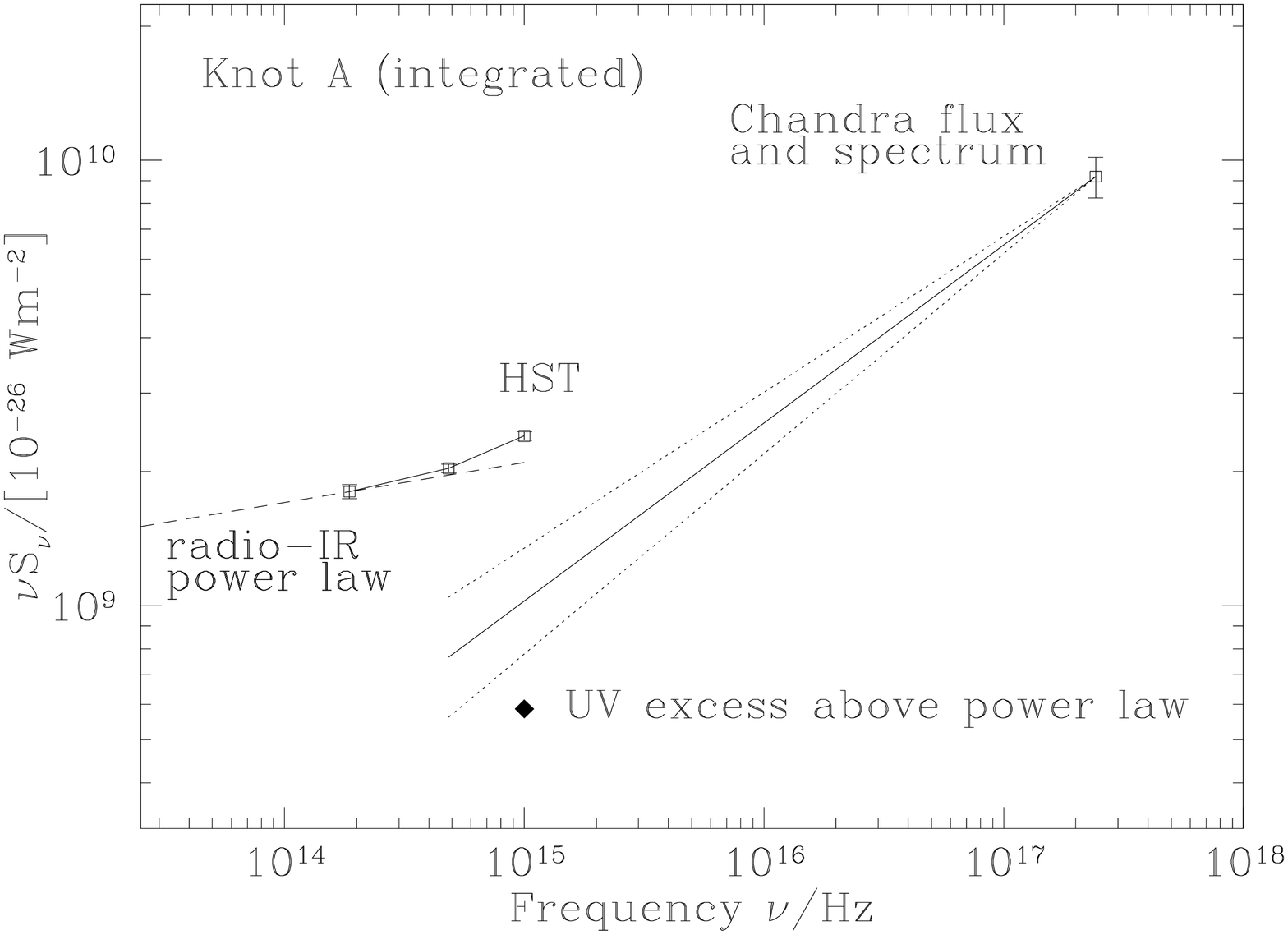}}\hfill\hfill
\resizebox{.46\hsize}{!}{\includegraphics[angle=270]{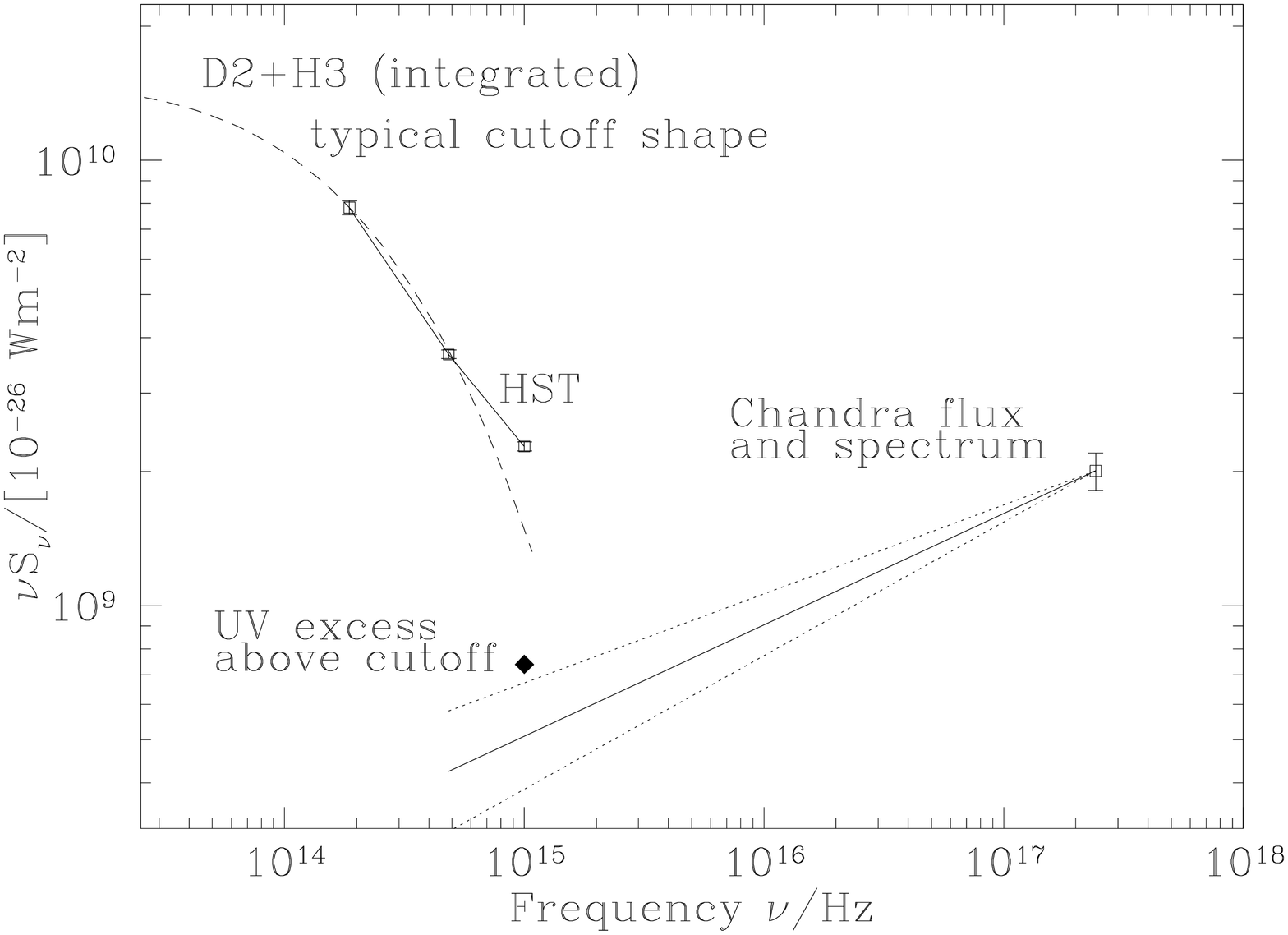}}\hfill
 \caption{Integrated high-frequency flux for regions A (left) and D2+H3 (right).
 Data are listed in Tab.\,\ref{t:sed}.  The UV flux in region A rises
 above a power-law extrapolation from the radio to the near-infrared.
 The UV flux in D2+H3 lies slightly above a power law extrapolation
 from near-infrared to optical, and clearly above a synchrotron cutoff
 (the shown spectrum merely illustrates the expected shape; see
 \S\ref{s:X-rays} for details about the determination of the UV
 excess). In both cases, the UV excess is consistent with the
 extrapolation of the X-ray flux at the X-ray spectral index
 \protect\citep[both determined by][]{Mareta01}.}  \label{f:Xrays}
\end{figure*}

The three high-frequency data points presented here are not sufficient
to fully disentangle contributions from two different processes.  We
therefore only consider whether the UV excess derived using our
spectral fits is consistent with an extrapolation of the X-ray flux
and spectral index.

The best current X-ray data have been presented by
\citet{Mareta01}. They have a spatial resolution of 0\farcs7. The
comparison is therefore carried out only for two regions of the jet
for which \citet{Mareta01} quote an X-ray flux and spectral index
(their Table 2) and for which the association with optical features is
clear: Region A1 and Region D2+H3 (beware of slightly different
nomenclature).  We sum the near-infrared, optical and ultraviolet flux
and the ultraviolet excess above the radio-infrared power law for A
and the fitted cutoff for D2/H3, respectively (Tab.\,\ref{t:sed}), and
compare these with a power-law extrapolation from X-rays.  As
evidenced by Fig.\,\ref{f:Xrays}, there is a surprisingly good match
in both regions between the UV excess determined here and the
extrapolation of the X-ray observations, suggesting a common origin.

We thus propose a two-component model for the emission of this jet: a
``low-energy'' synchrotron component and a second ``high-energy''
component, which accounts for the X-rays and is already noticeable at
ultraviolet wavelengths, either synchrotron emission \citep[the
observed X-ray spectral index is in the typical range for synchrotron
emission from jets;][]{hs_II} or beamed inverse Compton emission
(microwave photons are upscattered into the UV by electrons with
Lorentz factors of 5--10 for jet Lorentz factors of 20--5).  As
synchrotron spectra are necessarily convex in $\log S_\nu$ against
$\log\nu$, the observed high-frequency flattening implies the presence
of a second emission component in the UV even for different model
spectra from those assumed here.  Only the magnitude of the excess
depends on the spectral shape, and hence the proposed common origin of
UV excess and X-rays.  As evidenced by the small changes in the
radio-infrared spectral index, the shape of the ``low-energy''
electron energy distribution is kept constant within the entire jet by
some unknown physical mechanism.

With the present radio-optical data, no statements can be made about
the spectral shape of the ``high-energy'' component beyond the
plausibility argument presented here.  In order to confirm the reality
of the UV excess and to characterise its spectrum, it is both
necessary to constrain the run of the optical synchrotron spectrum
more accurately by further observations in the near-infrared and
optical, and to characterise the spectrum of the UV excess by
far-ultraviolet observations.  Deeper Chandra observations are needed
to establish the detailed X-ray morphology of the outermost part of
the jet and confirm the tentative detection of X-ray spectral index
changes \citetext{H.\ Marshall, {\it priv.comm.\/}}, and thus test our model of
a common origin for the ultraviolet excess and the jet's X-ray
emission.  Ideally, X-ray polarimetry would provide the most stringent
test for the X-ray emission mechanism beyond our spectral
studies. However, such a study will have to await the construction of
X-ray polarimeters with sufficient sensitivity and resolution, which
are currently in the development stage.

\begin{acknowledgements}
We acknowledge fruitful discussions with Herman Marshall.  We thank
the referee, Dan Harris, for rapid and thorough criticism.
\end{acknowledgements}

\end{document}